\documentclass[aps,prd,showpacs,preprintnumbers,twocolumn]{revtex4}
\usepackage{graphicx}
\usepackage{epsfig}
\usepackage{amsmath}
\usepackage{subfigure}
\def \beq{\begin{equation}}
\def \eeq{\end{equation}}
\def \beqa{\begin{eqnarray}}
\def \eeqa{\end{eqnarray}}
\begin{document}

\title{Dressed Polyakov loop and flavor dependent phase transitions}
\bigskip
\bigskip
\author{Fukun Xu$^{1}$}
\email{xufukun@mail.ihep.ac.cn}
\author{Hong Mao$^{1,2}$}
\email{hznu.mao@gmail.com}
\author{Tamal K. Mukherjee$^{1,3}$}
\email{mukherjee@mail.ihep.ac.cn}
\author{Mei Huang$^{1,3}$}
\email{huangm@mail.ihep.ac.cn} \affiliation{$^{1}$ Institute of High
Energy Physics, Chinese Academy of Sciences, Beijing, China }
\affiliation{$^{2}$ Department of Physics, Hangzhou Normal
University, Hangzhou 310036, China} \affiliation{$^{3}$ Theoretical
Physics Center for Science Facilities, Chinese Academy of Sciences,
Beijing, China}
\date{\today }
\bigskip

\begin{abstract}
The chiral condensate and dressed Polyakov loop at finite
temperature and density have been investigated in the framework of
$N_f=2+1$ Nambu--Jona-Lasinio (NJL) model with two degenerate $u,d$
quarks and one strange quark. In the case of explicit chiral
symmetry breaking with physical quark masses, it is found that the
phase transitions for light $u,d$ quarks and $s$ quark are
sequentially happened, and the separation between the transition
lines for different flavors becomes wider and wider with the
increase of baryon density. For each flavor, the pseudo-critical
temperatures for chiral condensate and dressed Polyakov loop differ
in a narrow transition range in the lower baryon density region, and
the two transitions coincide in the higher baryon density region.
\end{abstract}


\maketitle
\section{Introduction}
\label{intro}

QCD vacuum is characterized by spontaneous chiral symmetry breaking
and color confinement. The dynamical chiral symmetry breaking is due
to a non-vanishing quark anti-quark condensate, $\langle{\bar q}q
\rangle \simeq (250 {\rm MeV})^3$ in the vacuum, which induces the
presence of the light Nambu-Goldstone particles, the pions and kaons
in the hadron spectrum. The confinement represents that only
colorless states are observed in the spectrum, which is commonly
described by the linearly rising potential between two heavy quarks
at large distances, $V_{\bar{Q}Q}(R)=\sigma_s R$, where $\sigma_s
\simeq (425 {\rm MeV})^2$ is the string tension.

It is expected that chiral symmetry can be restored and color
degrees of freedom can be freed at high temperature and/or density.
The interplay between chiral and deconfinement phase transitions at
finite temperature and density are of continuous interests for
studying the QCD phase diagram
\cite{Polyakov:1978vu,'tHooft:1977hy,Casher:1979vw,Banks:1979yr,Hatta:2003ga,Mocsy:2003qw,Marhauser:2008fz,Braun:2007bx,Braun:2009gm}.
The chiral restoration is characterized by the restoration of chiral
symmetry and the deconfinement phase transition is characterized by
the breaking of center symmetry, which are only well defined in two
extreme quark mass limits, respectively. In the chiral limit when
the current quark mass is zero $m=0$, the chiral condensate
$\langle{\bar q}q \rangle$ is the order parameter for the chiral
phase transition. When the current quark mass goes to infinity
$m\rightarrow \infty$, QCD becomes pure gauge $SU(3)$ theory, which
is center symmetric in the vacuum, and the usually used order
parameter is the Polyakov loop expectation value $\langle P \rangle
$ \cite{Polyakov:1978vu}, which is related to the heavy quark free
energy. At zero density and chiral limit, lattice QCD results show
that the chiral and deconfinement phase transitions occur at the
same critical temperature, e.g, see Ref.
\cite{Kogut:1982rt,Fukugita:1986rr,Karsch:1994hm,Digal:2000ar,Digal:2002wn},
and also review papers \cite{Karsch:2001cy,Laermann:2003cv}. This
result is highly nontrivial because these two distinct phase
transitions involve different mechanisms at different energy scales.
It has been largely believed for a long time that chiral symmetry
restoration always coincides with deconfinement phase transition in
the whole $(T,\mu)$ plane.

However, for the case of finite physical quark mass, neither the
chiral condensate nor the Polyakov loop is a good order parameter.
For heavy quark, there is no dynamical chiral symmetry breaking
(e.g, see \cite{Chang-Liu}) thus no chiral restoration. On the other
hand, the linear potential description for confinement property is
not suitable for light quark system. In recent years, several
lattice groups have made much effort on investigating the chiral and
deconfinement phase transition temperatures with almost physical
quark masses, e.g, RBC-Bielefeld group \cite{Cheng:2006qk}, which
later merged with part of the MILC group \cite{Bernard:2004je} and
formed the hotQCD group \cite{Bazavov:2009zn,Cheng:2009zi}, and
Wuppertal-Budapest group
\cite{Aoki:2006we,Aoki:2006br,Aoki:2009sc,Fodor:2009ax,Borsanyi:2010bp}.
The result from the RBC-Bielefeld group in 2006 \cite{Cheng:2006qk}
found that the two pseudo-critical temperatures for $N_f=2+1$
coincide at $T_c=192(7)(4) {\rm MeV}$. The Wuppetal-Budapest group
found that for the case of $N_f=2+1$, there are three pseudocritical
temperatures, the transition temperature for chiral restoration of
$u,d$ quarks $T_c^{\chi(ud)}= 151(3)(3)~{\rm MeV}$, the transition
temperature for $s$ quark number susceptibility
$T_c^{s}=175(2)(4){\rm MeV}$ and the deconfinement transition
temperature $T_c^d=176(3)(4) {\rm MeV}$ from the Polyakov loop.
Recently, it is shown in \cite{Bazavov:2009mi,Petreczky:2010tf}, by
using an improved HISQ action, hotQCD collaboration results are
close to the Wuppetal-Budapest collaboration results.

The relation between the chiral and deconfinement phase transitions
has also attracted more interest recently in studying the phase
diagram at high baryon density region \cite{QCD-phase}. It is
conjectured in Ref. \cite{McLerran:2007qj} that in large $N_c$
limit, a confined but chiral symmetric phase, which is called
quarkyonic phase can exist in the high baryon density region. The
quarkyonic phase or chiral density wave state is due to the
quark-hole pairing near the Fermi surface. Nevertheless, it attracts
a lot of interests to study whether such a chiral symmetric but
confined phase can survive in real QCD phase diagram, and how it
competes with nuclear matter and the color superconducting phase
\cite{CSC}.

In the framework of QCD effective models, there is still no
dynamical model which can describe the chiral symmetry breaking and
confinement simultaneously. The main difficulty of effective QCD
model to include confinement mechanism lies in that it is difficult
to calculate the Polyakov loop analytically. Currently, the popular
models used to investigate the chiral and deconfinement phase
transitions are the Polyakov Nambu-Jona-Lasinio model (PNJL)
\cite{Fukushima:2003fw,Ratti:2005jh,Sasaki:2006ww,Ghosh:2006qh,Fu:2007xc,Zhang:2006gu,Fukushima:2008wg,Abuki:2008nm}
and Polyakov linear sigma model (PLSM)
\cite{Schaefer:2007pw,Mao:2009aq}. However, the shortcoming of these
models is that the temperature dependence of the Polyakov-loop
potential is put in by hand from lattice result, which cannot be
self-consistently extended to finite baryon density. Recently,
efforts have been made in Ref.\cite{Kondo:2010ts,Herbst:2010rf} to
derive a low-energy effective theory for confinement-deconfinement
and chiral-symmetry breaking/restoration.

Recent investigation revealed that quark propagator, heat kernels
can also act as an order parameter as they transform non trivially
under the center transformation related to deconfinement transition
\cite{Synatschke:2007bz,Synatschke:2008yt,Bilgici:2008ui}. The
exciting result is the behavior of spectral sum of the Dirac
operator under center transformation
\cite{Synatschke:2008yt,Gattringer:2006ci,Bruckmann:2006kx,Bilgici:2008qy}.
A new order parameter, called dressed Polyakov loop has been defined
which can be represented as a spectral sum of the Dirac operator
\cite{Bilgici:2008qy}. It has been found the infrared part of the
spectrum particularly plays a leading role in confinement
\cite{Synatschke:2008yt}. This result is encouraging since it gives
a hope to relate the chiral phase transition with the
confinement-deconfinement phase transition. The order parameter for
chiral phase transition is related to the spectral density of the
Dirac operator through Banks-Casher relation \cite{Banks:1979yr}.
Therefore, both the dressed Polyakov loop and the chiral condensate
are related to the spectral sum of the Dirac operator. Behavior of
the dressed Polyakov loop is mainly studied in the framework of
Lattice gauge theory
\cite{Bruckmann:2008br,Bilgici-thesis,Zhang:2010ui}. Apart from
that, studies based on Dyson-Schwinger equations
\cite{Fischer:2009wc,Fischer:2009gk,Fischer:2010fx} and PNJL model
\cite{Kashiwa:2009ki,Gatto:2010qs} have been carried out. In those
studies the role of dressed Polyakov loop as an order parameter is
discussed at zero chemical potential. The dressed Polyakov loop at
finite temperature and density has been investigated in the
two-flavor NJL model in Ref.\cite{Mukherjee:2010cp}. In this paper,
we show the phase diagram in the framework of three-flavor NJL model
by using the dressed Polyakov loop as an equivalent order parameter.

This paper is organized as follows: We introduce the dressed
Polyakov loop as an equivalent order parameter of confinement
deconfinement phase transition and the NJL model in Sec.
\ref{DPL-section}. Then in Sec.\ref{SU3-section}, we show the
results of three-flavor QCD phase diagram in $T-\mu$ plane in the
chiral limit and in the case of explicit chiral symmetry breaking,
respectively. At the end, we give the conclusion and discussion.

\section{Dressed Polyakov loop and the three-flavor NJL model}
\label{DPL-section}

We firstly introduce the dressed Polyakov loop. To do this we have
to consider a $U(1)$ valued boundary condition for the fermionic
fields in the temporal direction instead of the canonical choice of
anti-periodic boundary condition,
\begin{equation}
\psi(x,\beta) = e^{-i \phi} \psi(x,0),
\end{equation}
where $0\leq \phi < 2\pi $ is the phase angle and $\beta$ is the
inverse temperature.

Dual quark condensate $\Sigma_n$ is then defined by the Fourier
transform (w.r.t the phase $\phi$) of the general boundary condition
dependent quark condensate
\cite{Bilgici:2008qy,Bruckmann:2008br,Bilgici-thesis},
\begin{equation}
\Sigma_n =-{\int_0}^{2\pi} \frac{d\phi}{2\pi} e^{-i n \phi}
\langle\bar{\psi} \psi\rangle_\phi, \label{eq.dpl}
\end{equation}
where $n$ is the winding number.

Particular case of $n=1$ is called the dressed Polyakov loop which
transforms in the same way as the conventional thin Polyakov loop
under the center symmetry and hence is an order parameter for the
deconfinement transition
\cite{Bilgici:2008qy,Bruckmann:2008br,Bilgici-thesis}. It reduces to
the thin Polyakov loop and to the dual of the conventional chiral
condensate in infinite and zero quark mass limits respectively,
i.e., in the chiral limit $m \rightarrow 0$ we get the dual of the
conventional chiral condensate and in the $m\rightarrow \infty$
limit we have thin Polyakov loop
\cite{Bilgici:2008qy,Bruckmann:2008br,Bilgici-thesis}.

The Lagrangian of three-flavor NJL model \cite{NJL-report} is given
as
\begin{eqnarray}
  {\cal{L}} & = & \bar{\psi}(i\gamma^{\mu}\partial_{\mu}-m)\psi
    + G_s \sum_a \Big\{ (\bar{\psi}\tau_a\psi)^2 + (\bar{\psi}i\gamma_5\tau_a\psi)^2
    \Big\} \nonumber \\
   & - & K \Big\{ {\rm Det}_f[\bar{\psi}(1+\gamma_5)\psi] + {\rm Det}_f[\bar{\psi}(1-\gamma_5)\psi]
   \Big\}.
\label{Lagr-flavor-3}
\end{eqnarray}
Where $\psi=(u,d,s)^T$ denotes the transpose of the quark field, and
$m={\rm Diag}(m_u,m_d,m_s)$ is the corresponding mass matrix in the
flavor space. $\tau_a$ with $a=1,\cdots,N_f^2-1$ are the eight
Gell-Mann matrices, and ${\rm Det}_f$ means determinant in flavor
space. The last term is the standard form of the 't Hooft
interaction, which is invariant under $SU(3)_L\times SU(3)_R\times
U(1)_B$ symmetry, but breaks down the $U_A(1)$ symmetry.

The $\phi$ dependent thermodynamic potential in the mean field level
is given as following:
\begin{eqnarray}
\Omega_{\phi} & = & \sum_f{\Omega_{\phi,M_f}} +
2G_s\sum_f{\langle\sigma\rangle_{\phi,f}^2} \nonumber \\
    & - & 4K\langle\sigma\rangle_{\phi,u} \langle\sigma\rangle_{\phi,d}
   \langle\sigma\rangle_{\phi,s} \, ,
\label{thermal-pot-flavor-3} \end{eqnarray} with
\begin{eqnarray}
\Omega_{\phi,M_f} & = & -2N_c {\int_\Lambda} \frac{d^3 p}{(2\pi)^3}
    \Big[E_{p,f}+{1\over\beta}ln(1+e^{-\beta E_{p,f}^-}) \nonumber \\
    & + & {1\over\beta}ln(1+e^{-\beta
    E_{p,f}^+})\Big].
\end{eqnarray}
Where the sum is in the flavor space,
$E_{p,f}=\sqrt{p^2+M^2_{\phi,f}}$ and $E_{p,f}^{\pm}=E_{p,f} \pm
[\mu+i(\phi-\pi)T]$, with the constituent quark mass
\begin{equation}
M_{\phi,i}=m_i-4G_s\langle\sigma\rangle_{\phi,i}
+2K\langle\sigma\rangle_{\phi,j}\langle\sigma\rangle_{\phi,k},
\end{equation}
where $(i,j,k)$ is the quark flavor indices $(u,d,s)$, and
$\langle\sigma\rangle_{\phi,f}=\langle\bar{\psi}_f\psi_f\rangle_{\phi}$.
We will only consider isospin symmetric quark matter and define a
uniform chemical potential $\mu$ for u, d and s.

It is known that the NJL model lacks of confinement and the gluon
dynamics is encoded in a static coupling constant for four point
contact interaction. However, assuming that we can read the
information of confinement from the dual chiral condensate, it would
be interesting to see the behavior of the dressed Polyakov loop in a
scenario without any explicit mechanism for confinement.

The thermodynamic potential contains imaginary part. We take only
the real part of the potential and the imaginary phase factor is not
considered in this work. The mean field
$\langle\sigma\rangle_{\phi}$ is obtained by minimizing the
potential for each value of $\phi \in [0,2\pi)$ for fixed T and
$\mu$. The conventional chiral condensate is
$\langle\sigma\rangle_{\pi}=\langle\bar{\psi}\psi\rangle_{\pi}$. For
brevity from here onwards we will represent the conventional chiral
condensate as $<\sigma>$. The dressed Polyakov loop $\Sigma_{1}$ is
obtained by integrating over the angle.

\section{Phase diagram for three flavors }
\label{SU3-section}

We investigate phase transitions for two cases, i.e., in the chiral
limit and in the case of explicit chiral symmetry breaking with
physical quark mass, and the corresponding parameters are taken from
Ref.\cite{Rehberg-Klevansky1995} and
\cite{Buballa:2003qv,Abuki2010}:

\begin{table}[h]
\begin{tabular}{|c|c|c|c|c|}
\hline & $m_q$[MeV] &\ $m_s$[MeV]  &\ $\;\;G_s\Lambda^2\;\;$ & \
$\;\;K\Lambda^5\;\;$  \\ \hline \ chiral-limit \ & 0 & 0 & 1.926 &
12.36 \\ \hline
\ physical mass  \ & 5.5  & 140.7  & 1.835 & 12.36    \\
\hline
\end{tabular}
\caption{Two sets of parameters in 3-flavor NJL model: the current
quark mass $m_q$ for up and down quark and $m_s$ for strange quark,
coupling constants $G$ and $K$, with a spatial momentum cutoff
$\Lambda=602.3$ MeV.} \label{tab}
\end{table}

\subsection{Phase diagram in the chiral limit}

\begin{figure}[thbp]
\epsfxsize=7.5 cm \epsfysize=6.5cm \epsfbox{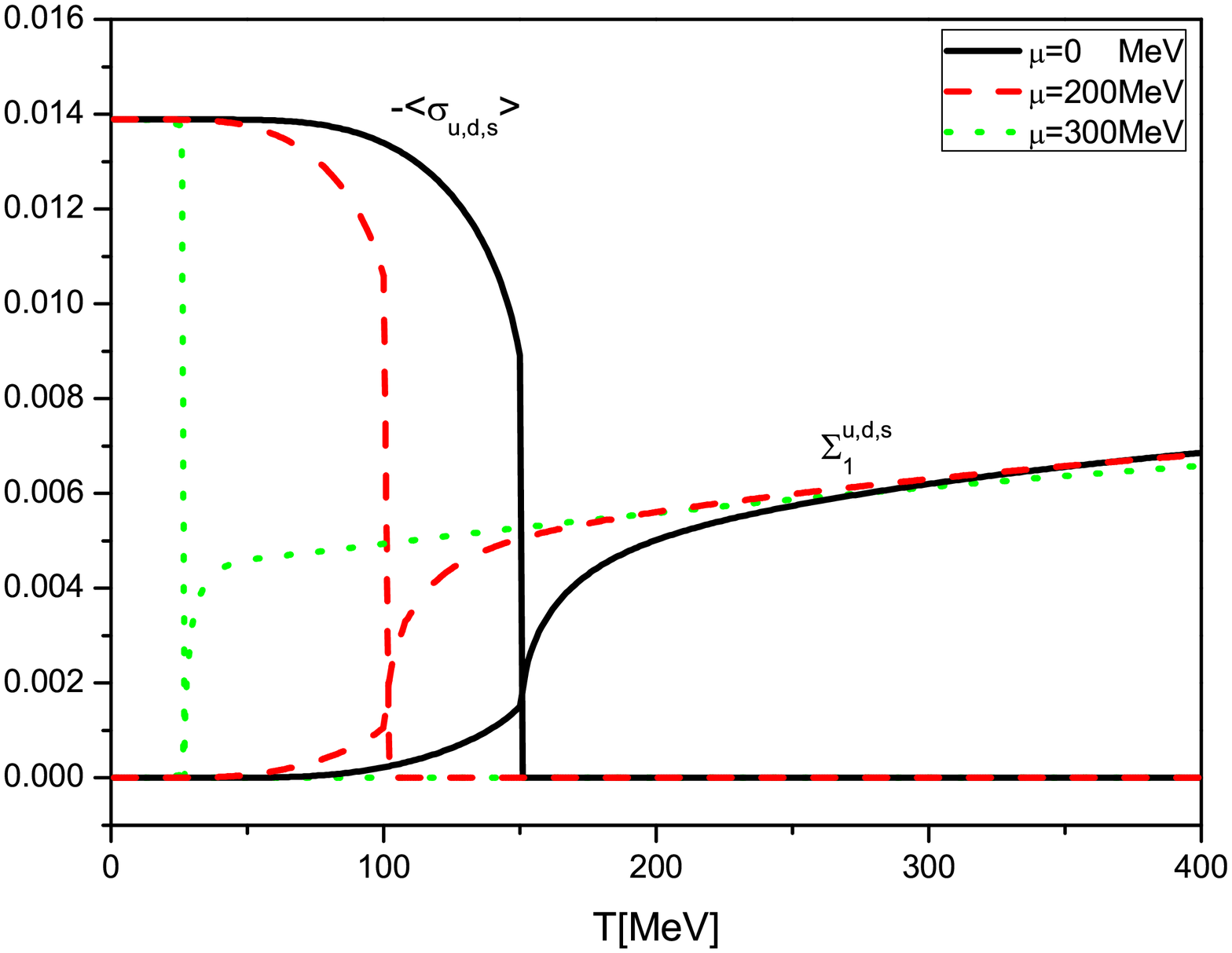} \caption{The
conventional chiral condensate $-\langle\sigma\rangle_{u,d,s}$ and
the dressed Polyakov loop $\Sigma_1^{u,d,s}$ of $u,d,s$ quarks as
functions of temperature for different values of the chemical
potentials. Here, $-\langle\sigma\rangle$ and $\Sigma_1$ both are
measured in $[{\rm GeV}^3]$.  } \label{fig-uds-chiral-limit}
\end{figure}

\begin{figure}[thbp]
\epsfxsize=7.5 cm \epsfysize=6.5cm \epsfbox{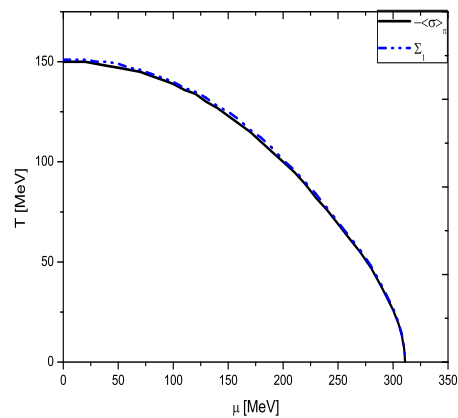}
\caption{Three-flavor phase diagram in the $T-\mu$ plane for the
case of chiral limit. The solid line is the critical line for
$\Sigma_1$, and the dashed line is the critical line for
conventional chiral phase transition. } \label{fig-SU3-chiral-limit}
\end{figure}

We firstly consider the case of chiral limit, i.e. $m_u=m_d=m_s=0$.
In Fig. \ref{fig-uds-chiral-limit}, we show the behavior of the
conventional chiral condensate $-\langle\sigma\rangle$ and the
corresponding dressed Polyakov loop $\Sigma_1$ for $u,d$ and $s$
quarks at different chemical potentials as functions of temperature.
For both order parameters, it is observed there are three
temperature regions for $-\langle\sigma\rangle$ and $\Sigma_1$. For
$-\langle\sigma\rangle$, at smaller temperatures it remains constant
at a value corresponding to the value of the conventional chiral
condensate in the vacuum, then it drops to zero at the critical
temperature $T_c$, and eventually keeps zero above the critical
temperature. The critical temperature decreases with the increase of
the chemical potential. It is noticed that, in order to guide eyes,
we have connected the two end-points of the order parameter at the
jump.

On the other hand the behavior for the dressed Polyakov loop is just
the opposite. It remains zero for small temperatures and then jumps
at the critical temperature, and finally saturates to a high value
which varies very slowly with temperatures. The almost zero value of
$\Sigma_1$ for small temperatures is due to the fact that the $U(1)$
boundary condition dependent general quark condensate nearly does
not vary with the angle $\phi$ for small temperatures (see
Eq.~\ref{eq.dpl}).

It is seen that the phase transitions for chiral restoration and
dressed Polyakov loop are of 1st order in the whole $T-\mu$ plane.
For two-flavor case, it was found these two phase transitions are of
second order. The $N_f$ dependent result is in agreement with the
results given by Pisarski and Wilczek in Ref.
\cite{Pisarski:1983ms}. The first order phase transition in
three-flavor case is due to the 't Hooft interaction in Eq.
(\ref{Lagr-flavor-3}), which contributes a cubic term in the
thermodynamical potential in Eq. (\ref{thermal-pot-flavor-3}).

Fig. \ref{fig-SU3-chiral-limit} shows the phase diagram of
three-flavor in the chiral limit. We find almost exact matching for
the transition temperatures calculated from these two quantities in
the whole $T-\mu$ plane.

\subsection{Phase diagram with physical quark mass}

For the case of finite quark mass $m_u=m_d=5.5 {\rm MeV}$ and
$m_s=140.7{\rm MeV}$, we have chosen the model parameters of $G_s
\Lambda^2=1.835$, $K \Lambda^5=12.36$ with $\Lambda= 602.3 {\rm
MeV}$ as in Ref. \cite{Rehberg-Klevansky1995} to fit $m_{\pi}=135.0
{\rm MeV}$, $f_{\pi}=92.4 {\rm MeV}$, $m_K=497.7 {\rm MeV}$ and
$m_{\eta'}=957.8 {\rm MeV}$.

\begin{figure}[thbp]
\epsfxsize=7.5 cm \epsfysize=6.5cm \epsfbox{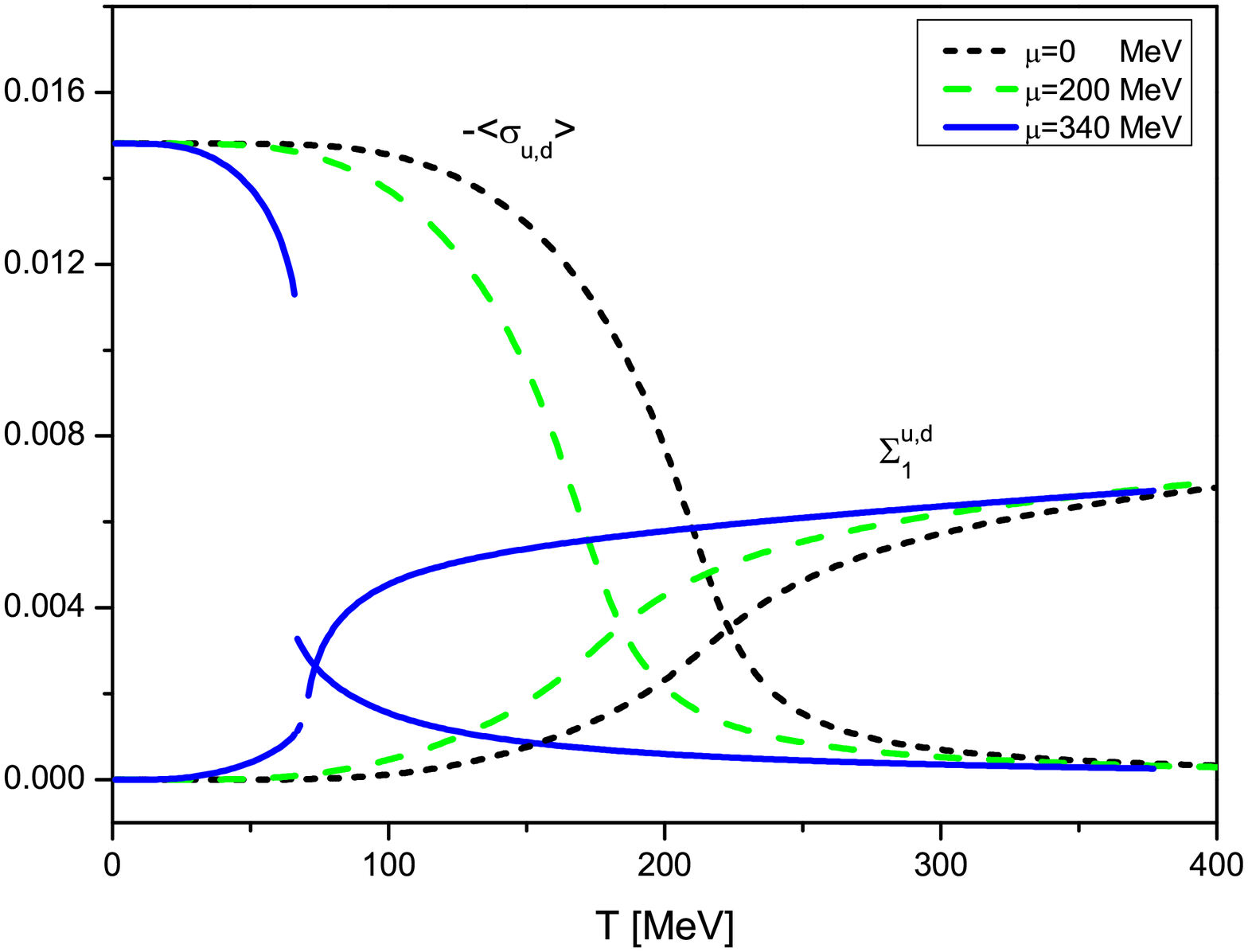} \caption{The
conventional chiral condensate $-\langle\sigma\rangle_{u,d}$ and the
dressed Polyakov loop $\Sigma_1^{u,d}$ of $u,d$ quarks as functions
of temperature for different values of the chemical potentials.
Here, $-\langle\sigma\rangle$ and $\Sigma_1$ both are measured in
$[{\rm GeV}^3]$. } \label{fig-ud-mu}
\end{figure}

\begin{figure}[thbp]
\epsfxsize=7.5 cm \epsfysize=6.5cm \epsfbox{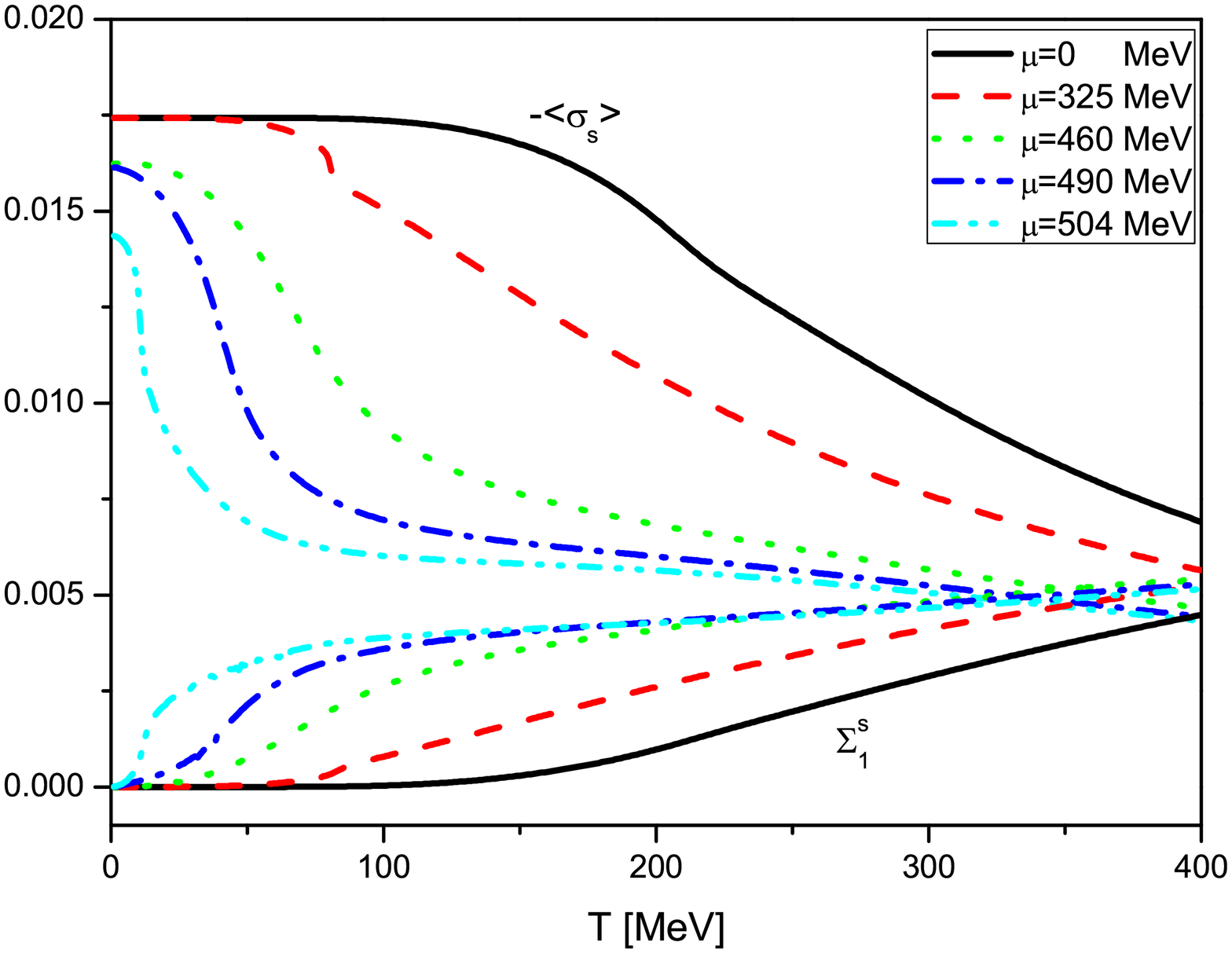} \caption{The
conventional chiral condensate $-\langle\sigma\rangle_s$ and the
dressed Polyakov loop $\Sigma_1^s$ of $s$ quark as functions of
temperature for different values of the chemical potentials. Here,
$-\langle\sigma\rangle$ and $\Sigma_1$ both are measured in $[{\rm
GeV}^3]$. } \label{fig-s-mu}
\end{figure}

In Fig. \ref{fig-ud-mu} and \ref{fig-s-mu}, we show the behavior of
the conventional chiral condensate $-\langle\sigma\rangle$ and the
dressed Polyakov loop $\Sigma_1$ at different chemical potentials as
functions of temperature for $u,d$ and $s$ quarks, respectively.

For both cases, it is observed that there are three temperature
regions for $-\langle\sigma\rangle$ and $\Sigma_1$. For
$-\langle\sigma\rangle$, at smaller temperatures it remains constant
at a value corresponding to the value of the conventional chiral
condensate in the vacuum, then it rapidly decreases in a small
window of temperature and eventually almost saturates to a lower
value. The decreasing occurs at different temperatures for different
values of the chemical potentials. On the other hand the behavior
for the dressed Polyakov loop is just the opposite. It remains
almost zero for small temperatures and then rises rapidly, finally
saturates to a high value which varies very slowly with
temperatures. The almost zero value of $\Sigma_1$ for small
temperatures is due to the fact that the $U(1)$ boundary condition
dependent general quark condensate nearly does not vary with the
angle $\phi$ for small temperatures (see Eq.~\ref{eq.dpl}).

\begin{figure}[thbp]
\epsfxsize=7.5 cm \epsfysize=6.5cm \epsfbox{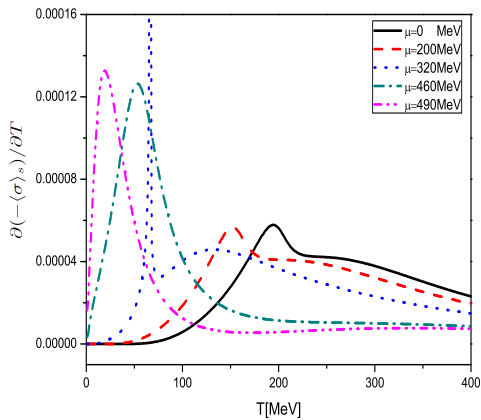} \caption{The
derivative of strange chiral condensate
$\partial(-{\langle\sigma\rangle_s)}/{\partial T}$ as functions of
$T$ for different values of $\mu$. } \label{fig-s-derive}
\end{figure}

The critical temperature for a real phase transition or the
pseudo-critical temperature for a crossover is extracted from the
susceptibility of the order parameter or the temperature derivative
of the order parameter. For example, for chiral phase transition of
strange quark, the (pseudo)critical temperature is extracted from
the temperature derivative $\partial (-<\sigma_s>)/(\partial T)$.
This quantity describes how fast the order parameter changes with
temperature. Normally the critical temperature corresponds to the
fastest change of the order parameter, and the temperature
derivative of the order parameter shows a peak at the critical
point. However, there are some subtleties to determine the
pseudo-critical temperature for the chiral restoration of the
strange quark. We show how we determine the pseudo-critical
temperature of the crossover by using Fig.\ref{fig-s-derive}, which
is the temperature derivative of the chiral condensate of the
strange quark corresponding to Fig.\ref{fig-s-mu}.

For $\mu=0$, from Fig.\ref{fig-s-derive} one can observe that the
temperature derivative of the strange quark condensate shows a peak
at $T= 196 {\rm MeV}$, correspondingly, from Fig.\ref{fig-s-mu}, one
can see that the strange quark condensate changes fast at $T= 196
{\rm MeV}$, which is the critical temperature for chiral phase
transition of the $u,d$ quarks at zero chemical potential. However,
the value of the strange quark condensate at $T_{c,\chi}^{u,d}= 196
{\rm MeV}$ is still around its vacuum value, one cannot locate the
pseudo-critical temperature of the strange quark at
$T_{c,\chi}^{u,d}= 196 {\rm MeV}$ even though there is a peak for
the temperature derivative of the strange quark condensate. The
reasonable explanation of the fast change of the strange quark
condensate at $T_{c,\chi}^{u,d}= 196 {\rm MeV}$ is that the strange
quark feels the chiral phase transition of $u,d$ quarks due to the
flavor mixing effect. For $\mu=0$, from Fig.\ref{fig-s-derive} one
can also observe a bump region of the temperature derivative of the
strange quark condensate around $T=250 {\rm MeV}$, however, there is
no obvious peak shown up. Therefore, we cannot extract an explicit
pseudo-critical temperature from the chiral phase transition of the
strange quark. Correspondingly, we find that the strange quark
condensate at $\mu=0$ changes smoothly with temperature.

The temperature derivative of the strange quark condensate at
$\mu=200 {\rm MeV}$ is similar to the case at $\mu=0$. The only
difference is that the peak moves to a lower temperature. The small
jump at the large strange chiral condensate region is induced by the
$u,d$ quark chiral phase transition. It cannot be regarded as the
phase transition for strange quark even though it corresponds to a
peak of the strange chiral susceptibility, because the order
parameter does not change so much comparing with its vacuum value.
It should still be regarded as in the chiral symmetry breaking
phase. At $\mu=320 {\rm MeV}$, it is seen from
Fig.\ref{fig-s-derive} that the left peak develops to a sharp peak
at $T_{c,\chi}^{u,d}$, and an obvious peak shows up in the right
bump region. Therefore, one can extract the pseudo-critical
temperature for the chiral phase transition of the strange quark.
For higher chemical potential, e.g, $\mu=460 {\rm MeV}$ or $\mu=490
{\rm MeV}$, because $u,d$ quarks are already in chiral symmetric
phase, there is only one peak shows up for the temperature
derivative of the strange quark condensate in Fig.
\ref{fig-s-derive}, and the location of the peak gives the
pseudo-critical temperature of the phase transition.

\begin{figure}[thbp]
\epsfxsize=7.5 cm \epsfysize=6.5cm \epsfbox{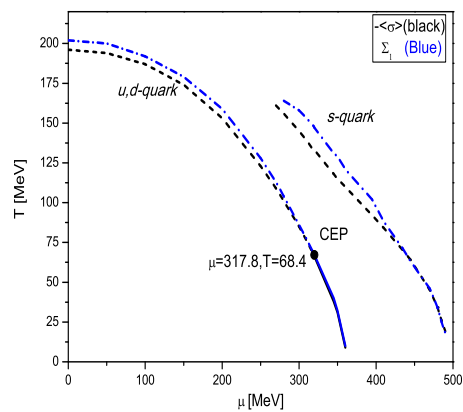}
\caption{Three-flavor phase diagram in the $T-\mu$ plane for the
case of $m_u=m_d=5{\rm MeV}$ and $m_s=140.7{\rm MeV}$. The
dash-dotted lines are the critical line for $\Sigma_1$, and the
dashed lines are the critical line for conventional chiral phase
transition in the region of crossover. The solid lines indicates the
1st order phase transitions, and the solid circle indicates the
critical end points for chiral phase transitions of $u,d$ quarks.}
\label{fig-SU3}
\end{figure}

As we have discussed in detail above, one has to combine the
information from the order parameter itself as well as the
temperature derivative of the order parameter in order to determine
the pseudo-critical temperature of the crossover. This method is
also used to determine the dressed Polyakov loop of the strange
quark. The critical and pseudo-critical temperatures extracted from
the temperature derivative of the order parameters are shown in Fig.
\ref{fig-SU3}. It is found that the the chiral and deconfinement
phase transitions are flavor dependent.

At low baryon chemical potential region when $\mu<270 {\rm MeV}$,
for light flavors, i.e. for $u,d$ quarks, we observe from Fig.
\ref{fig-ud-mu} that the conventional chiral condensate and the
dressed Polyakov loop change rapidly with the increase of
temperature. From the temperature derivative of the order parameters
of the chiral condensate and dressed Polyakov loop, we can obtain
two separate pseudo-critical temperatures $T_c^{\chi}$ and
$T_c^{\cal D}$ for fixed $\mu$, and we find $T_c^{\chi}$ is always
smaller than $T_c^{\cal D}$.

However, in the chemical potential region when $\mu<270 {\rm MeV}$,
for $s$ quark, from Fig. \ref{fig-s-mu} we can see that the
conventional chiral condensate and dressed Polyakov loop change
smoothly with the increase of temperature. From the temperature
derivative of the order parameters, one cannot extract the values of
the pseudo-critical temperatures as already discussed. Therefore, in
Fig. \ref{fig-SU3} of the three-flavor phase diagram, we can read
that in the region around $0<\mu<270 {\rm MeV}$, the phase
transitions for $u,d$ are crossover, and different order parameters
have different pseudo-critical temperatures. The $s$ flavor
experiences a rapid crossover, and no pseudo-critical temperatures
can be extracted from the order parameters. From the lattice results
in Ref.\cite{Aoki:2009sc} at zero chemical potential, there is also
no pseudo-critical temperature for the order parameter of strange
quark's chiral condensate.

At higher baryon chemical potential region, it is observed from
Fig.\ref{fig-ud-mu} that the conventional chiral condensate and the
dressed Polyakov loop change sharply with the increase of
temperature. From the temperature derivative of the order
parameters, we find that the phase transitions are of first order,
and the critical temperatures for chiral and dressed Polyakov loop
coincide with each other around CEP.

For $s$ quark, from Fig. \ref{fig-s-mu} we can see that when the
chemical potential becomes higher and higher, the conventional
chiral condensate and dressed Polyakov loop change more rapidly with
the increase of temperature. The temperature derivative of the order
parameters give separate values of the pseudo-critical temperatures
in the region $270 <\mu< 450 {\rm MeV}$, and the two pseudo-critical
temperatures merge in the region of $\mu>450 {\rm MeV}$.

From Fig. \ref{fig-SU3} of the three-flavor phase diagram, we can
read the critical end point for $u,d$ flavors lies at
$(T_{CEP}^{u,d},\mu_{CEP}^{u,d})=(68.4{\rm MeV},317.8{\rm MeV})$,
which is different from the results in Ref.\cite{Mukherjee:2010cp}
for pure two-flavor NJL model. The difference comes from: 1)
different model parameters have been used, 2) the coupling of $s$
quark to $u,d$ quark contributes one extra term in the
thermodynamical potential comparing with the pure two-flavor case.
The location of CEP in this work is in good agreement with that in
Ref. \cite{costa-NJL}.

\begin{figure}[thbp] \epsfxsize=7.5 cm
\epsfysize=6.5cm \epsfbox{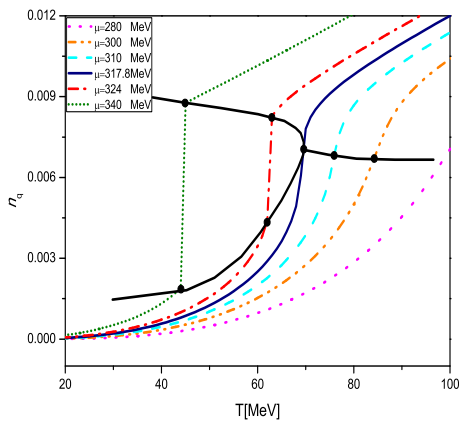} \caption{The quark number
density $n_q$ as functions of the temperature for different chemical
potentials, and $n_q$ is in unit of $[{\rm GeV}^3]$.}
\label{fig-chiq}
\end{figure}

\begin{figure}[thbp]
\epsfxsize=7.5 cm \epsfysize=6.5cm \epsfbox{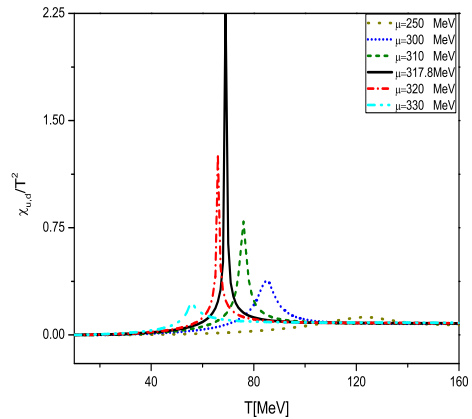} \caption{The
number susceptibility $\chi_q/T^2$ as functions of the temperature
for different chemical potentials.} \label{fig-CEP}
\end{figure}

In Fig.\ref{fig-chiq} and Fig.\ref{fig-CEP}, we show the details of
locating the CEP. In the first order phase region, there are two
branches of number densities, i.e, for fixed chemical potential, the
number density $n_q=-\frac{\partial\Omega}{\partial\mu}$ has a jump
at the transition temperature. The two branches of number densities
merge at the CEP. This feature is shown in Fig.\ref{fig-chiq}. At
the CEP, the phase transition is of second order and this is
indicated by the divergent behavior of the number susceptibility. We
show the number susceptibility
$\chi_q=-\frac{\partial^2\Omega}{\partial\mu^2}$ as functions of the
temperature in Fig.\ref{fig-CEP}. It is clearly seen that $\chi_q$
develops a sharp peak at CEP.

\section{Conclusion and discussion}

\begin{figure}[thbp]
\epsfxsize=7.5 cm \epsfysize=6.5cm \epsfbox{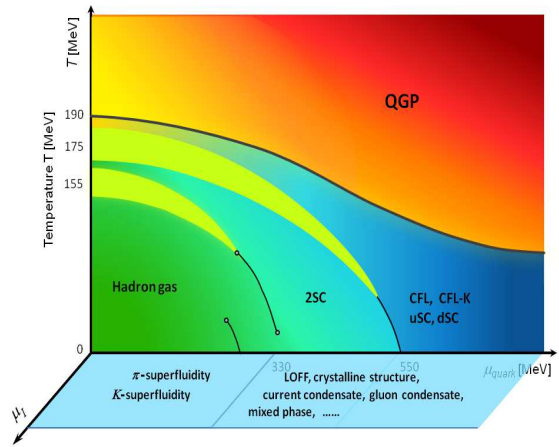}
\caption{Conjectured 3D QCD phase diagram at finite temperature $T$,
quark chemical potential $\mu_q$ and isospin chemical potential
$\mu_I$. } \label{fig-conjecture-phase}
\end{figure}

We investigate the chiral condensate and the dressed Polyakov loop
or dual chiral condensate at finite temperature and density in the
three-flavor Nambu--Jona-Lasinio model. It is found that in the
chiral limit, the phase transitions are of 1st order and the
critical temperature for chiral phase transition coincides with that
of the dressed Polyakov loop. In the case of explicit chiral
symmetry breaking, it is found that the phase transitions are flavor
dependent, and there is a phase transition range for each flavor.
The transition range of $s$ quark is located at higher temperature
and higher baryon density than that of $u,d$ quarks. At low baryon
density region, it is found that the transition range of $u,d$
quarks are not separated too much from that of the $s$ quark,
however, the separation of the transition ranges for $u,d$ quarks
and $s$ quark become wider and wider with the increase of the
chemical potential.

For light $u,d$ quarks, the pseudo-critical temperature for chiral
transition $T_c^{\chi}$ is smaller than that of the dressed Polyakov
loop $T_c^{{\cal D}}$ in the low baryon density region where the
transition is a crossover, and these two phase transitions coincide
in the 1st order phase transition region at high baryon density. For
$s$ quark, both transitions are of smooth crossover at low baryon
density, and becomes rapid crossover at moderate baryon density
region where the pseudo-critical temperatures for the chiral
condensate and the dressed Polyakov loop are separated, then at
enough high baryon density, these two transitions coincide with each
other.

Our results are based on the NJL model, where the gluon dynamics is
encoded in a static coupling constant for four point contact
interaction, a quantitative comparison will not match with lattice
results. However, we believe the scenario of the sequential phase
transitions is physically correct.

Till now, there are six quark flavors observed in experiment. These
six flavors cover a very wide energy scale, from several ${\rm MeV}$
to several hundred ${\rm GeV}$. Only light quarks experience
dynamically chiral symmetry breaking in the vacuum, and chiral phase
transition in high temperature and density. However, there is no
good order parameters to describe the deconfinement phase transition
of light quarks. The conventional Polyakov loop is a good order
parameter for confinement deconfinement phase transition in the
limit of infinity heavy quark mass, and has the interpretation of
the free energy of an infinity heavy quark. In analogy to that we
can regard the dressed Polyakov loop as an order parameter for
confinement deconfinement phase transition for a quark with mass
$m$, and interpret the dressed Polyakov loop as the free energy of a
quark with any mass $m$ \cite{Zhang:2010ui}. Therefore, in
principle, each flavor can have different critical temperatures for
deconfinement phase transition. Lattice results already reflect such
properties at zero chemical potential, e.g. the pseudo-critical
temperatures for order parameters of $u,d$ quarks, $s$ quark and the
Polyakov loop are different, and the the pseudo-critical temperature
is higher for heavier quark mass.

It is natural to understand that the separation of the phase
transition range for different flavors becomes wider and wider with
the increase of the chemical potential. Lattice result at zero
chemical potential gives that the pseudo-critical temperature for
$u,d$ quarks is around $155 {\rm MeV}$, and for $s$ quark is around
$175 {\rm MeV}$. The difference is around $20 {\rm MeV}$. However,
at zero temperature, the $u,d$ quarks restores chiral symmetry at
the chemical potential around their vacuum constituent masses, i.e.
$\mu_c^{u,d}\sim M_{u,d}\sim 330 {\rm MeV}$, and the $s$ quark
restores chiral symmetry at the chemical potential around
$\mu_c^{s}\sim M_{s}\sim 550 {\rm MeV}$. The difference is around
$200 {\rm MeV}$.

Based on above analysis, in Fig. \ref{fig-conjecture-phase}, we show
our conjectured 3 dimension (3D) QCD phase diagram for finite
temperature $T$, quark chemical potential $\mu_q$ and isospin
chemical potential $\mu_I$.

In the plane of $(\mu, T)$, each flavor has its own transition
range. The transition range is wider in the low baryon density, and
becomes narrower and narrower with the increase of the chemical
potential, and eventually merge at higher chemical potential. By
using the lattice results at zero density, we identify the phase
transition range around $155 {\rm MeV}$ for $u,d$ quark, $175 {\rm
MeV}$ for $s$ quark, and $190 {\rm MeV}$ for heavy flavor. The upper
solid line is for the Polyakov loop, which does not change so much
with the increase of baryon density. This result agrees with that in
any Polyakov loop NJL model and Polyakov loop linear sigma model.
Due to the flavor dependent phase transitions, we naturally expect
the color superconducting phase for two-flavor quark system and
three-flavor quark system in different baryon density regions
\cite{CSC}. Due to the finite mass of strange quark, the
three-flavor color superconducting phase can be in the color flavor
locking (CFL) phase\cite{CFL}, CFL-kaon condensate phase
(CFL-K)\cite{CFL-K}, or uSC/dSC phase \cite{udSC}.

When isospin asymmetry is considered, the phase diagram becomes much
more complicated. At low baryon density region, there will be pion
superfluidity and kaon superfluidity phases \cite{He:2005nk}. In the
color superconducting phase, because isospin asymmetry induces
mismatch between the pairing quarks, there will appear unstable
gapless excitations \cite{g2SC,gCFL} when charge neutrality
condition is considered. It has been vastly discussed in many
literatures about the true ground state of the charge neutral
two-flavor and three-flavor cold quark matter, e.g, the
Larkin-Ovchinnikov-Fulde-Ferrell (LOFF) sate or other crystalline
structure \cite{LOFF}, the gluon condensate state \cite{GC}, the
current generation state \cite{current}, and so on. The detailed
analysis given in Ref. \cite{Higgs-current} show that in the gapless
color superconducting phase, both the phase part and magnitude part
of the order parameter will develop instabilities. The phase part
develops into the chromomagnetic instability, which induces the
plane-wave state; The magnitude part develops the Sarma instability
and Higgs instability, the Sarma instability can be competed with
charge neutrality condition. If the Higgs instability cannot be
cured by the electric or color Coulomb interaction, it will induce
the inhomogeneous state.

\vskip 1cm \noindent

{\bf Acknowledgments}: The work of M.H. is supported by CAS program
"Outstanding young scientists abroad brought-in", and NSFC under the
number of 10735040 and 10875134, and K.C.Wong Education Foundation,
Hong Kong. The work of H.M is supported by NSFC 10904029 and the
Natural Science Foundation of Zhejiang Province under grant no
Y7080056.


\begin{thebibliography}{9}

%%%% Polyakov:1978vu,
%%%%%%% Confinement versus chiral symmetry breaking %%%%%%
%%% Hatta:2003ga,Mocsy:2003qw,
%%%%Polyakov:1978vu,'tHooft:1977hy,Casher:1979vw,Banks:1979yr
%%% Hatta:2003ga,Mocsy:2003qw,McLerran:2007qj,Fukushima:2008wg
\bibitem{Polyakov:1978vu}
  A.~M.~Polyakov,
  %``Thermal Properties Of Gauge Fields And Quark Liberation,''
  Phys.\ Lett.\  B {\bf 72}, 477 (1978).
  %%CITATION = PHLTA,B72,477;%%

\bibitem{'tHooft:1977hy}
  G.~'t Hooft,
  %``On The Phase Transition Towards Permanent Quark Confinement,''
  Nucl.\ Phys.\  B {\bf 138}, 1 (1978).
  %%CITATION = NUPHA,B138,1;%%

\bibitem{Casher:1979vw}
  A.~Casher,
  %``Chiral Symmetry Breaking In Quark Confining Theories,''
  Phys.\ Lett.\  B {\bf 83}, 395 (1979).
  %%CITATION = PHLTA,B83,395;%%

\bibitem{Banks:1979yr}
  T.~Banks and A.~Casher,
  %``Chiral Symmetry Breaking In Confining Theories,''
  Nucl.\ Phys.\  B {\bf 169}, 103 (1980).
  %%CITATION = NUPHA,B169,103;%%

\bibitem{Hatta:2003ga}
  Y.~Hatta and K.~Fukushima,
  %``Linking the chiral and deconfinement phase transitions,''
  Phys.\ Rev.\  D {\bf 69}, 097502 (2004).
%  [arXiv:hep-ph/0307068].
  %%CITATION = PHRVA,D69,097502;%%

\bibitem{Mocsy:2003qw}
  A.~Mocsy, F.~Sannino and K.~Tuominen,
  %``Confinement versus Chiral Symmetry,''
  Phys.\ Rev.\ Lett.\  {\bf 92}, 182302 (2004).
  %[arXiv:hep-ph/0308135].
  %%CITATION = PRLTA,92,182302;%%

\bibitem{Marhauser:2008fz}
  F.~Marhauser and J.~M.~Pawlowski,
  %``Confinement in Polyakov Gauge,''
  arXiv:0812.1144 [hep-ph].
  %%CITATION = ARXIV:0812.1144;%%

\bibitem{Braun:2007bx}
  J.~Braun, H.~Gies and J.~M.~Pawlowski,
  %``Quark Confinement from Color Confinement,''
  Phys.\ Lett.\  B {\bf 684}, 262 (2010).
%  [arXiv:0708.2413 [hep-th]].
  %%CITATION = PHLTA,B684,262;%%
%%%
\bibitem{Braun:2009gm}
  J.~Braun, L.~M.~Haas, F.~Marhauser and J.~M.~Pawlowski,
  %``On the relation of quark confinement and chiral symmetry breaking,''
  arXiv:0908.0008 [hep-ph].
  %%CITATION = ARXIV:0908.0008;%%

\bibitem{QCD-phase}
T.~Hatsuda and K.~Maeda,
  %``Quantum Phase Transitions in Dense QCD,''
  arXiv:0912.1437 [hep-ph];
  %%CITATION = ARXIV:0912.1437;%%
K.~Fukushima and T.~Hatsuda,
  %``The phase diagram of dense QCD,''
  Rept.\ Prog.\ Phys.\  {\bf 74}, 014001 (2011);
 % [arXiv:1005.4814 [hep-ph]].
  %%CITATION = RPPHA,74,014001;%%
J.~M.~Pawlowski,
  %``The QCD phase diagram: Results and challenges,''
  arXiv:1012.5075 [hep-ph].
  %%CITATION = ARXIV:1012.5075;%%

\bibitem{McLerran:2007qj}
  L.~McLerran and R.~D.~Pisarski,
  %``Phases of Cold, Dense Quarks at Large N_c,''
  Nucl.\ Phys.\  A {\bf 796}, 83 (2007).
 % [arXiv:0706.2191 [hep-ph]].
  %%CITATION = NUPHA,A796,83;%%

\bibitem{Chang-Liu}
  L.~Chang, Y.~X.~Liu, M.~S.~Bhagwat, C.~D.~Roberts and S.~V.~Wright,
  %``Dynamical chiral symmetry breaking and a critical mass,''
  Phys.\ Rev.\  C {\bf 75}, 015201 (2007)
  [arXiv:nucl-th/0605058].
  %%CITATION = PHRVA,C75,015201;%%


%%%%%% CSC and Quarkyonic phase %%%%%%%%
\bibitem{CSC}K.~Rajagopal and F.~Wilczek,
%``The condensed matter physics of QCD,''
hep-ph/0011333;
%%CITATION = HEP-PH 0011333.%%
D.K.~Hong,
%``Aspects of color superconductivity,''
Acta Phys.\ Polon.\ B {\bf 32}, 1253 (2001);
%%CITATION = HEP-PH 0101025;%%
M.~Alford,
%``Color superconducting quark matter,''
Ann.\ Rev.\ Nucl.\ Part.\ Sci.\  {\bf 51}, 131 (2001);
%%CITATION = HEP-PH 0102047;%%
G.~Nardulli,
%``Effective description of QCD at very high densities,''
Riv.\ Nuovo Cim.\  {\bf 25N3}, 1 (2002);
%%[ hep-ph/0202037].
%%CITATION = HEP-PH 0202037;%%
T.~Sch{\"a}fer,
%``Quark matter,''
hep-ph/0304281;
%%CITATION = HEP-PH 0304281;%%
H.C.~Ren,
%``Color superconductivity of QCD at high baryon density,''
hep-ph/0404074;
%%CITATION = HEP-PH 0404074;%%
M.~Huang,
  %``Color superconductivity at moderate baryon density,''
  Int.\ J.\ Mod.\ Phys.\  E {\bf 14}, 675 (2005);
%  [arXiv:hep-ph/0409167].
 I.~A.~Shovkovy,
  %``Two lectures on color superconductivity,''
  Found.\ Phys.\  {\bf 35}, 1309 (2005);
%  [arXiv:nucl-th/0410091].
  %%CITATION = FNDPA,35,1309;%%
M.~G.~Alford, A.~Schmitt, K.~Rajagopal and T.~Schafer,
  %``Color superconductivity in dense quark matter,''
  Rev.\ Mod.\ Phys.\  {\bf 80}, 1455 (2008);
%  [arXiv:0709.4635 [hep-ph]].
  %%CITATION = RMPHA,80,1455;%
Q.~Wang,
  %``Some aspects of color superconductivity: an introduction,''
arXiv:0912.2485 [nucl-th];
  %%CITATION = ARXIV:0912.2485;%%

\bibitem{Panero:2009tv}
  M.~Panero,
  %``Thermodynamics of the QCD plasma and the large-N limit,''
  Phys.\ Rev.\ Lett.\  {\bf 103}, 232001 (2009).
 % [arXiv:0907.3719 [hep-lat]].
  %%CITATION = PRLTA,103,232001;%%

%%%% LQCD %%%%%
%%Kogut:1982rt,Digal:2002wn,

%%%chiral limit %%%
%%%%LQCD-coincide Kogut:1982rt,Fukugita:1986rr,Karsch:1994hm,Digal:2002wn
%%%% Karsch:2001cy,Laermann:2003cv
\bibitem{Kogut:1982rt}
  J.~B.~Kogut, M.~Stone, H.~W.~Wyld, W.~R.~Gibbs, J.~Shigemitsu, S.~H.~Shenker and D.~K.~Sinclair,
  %``Deconfinement And Chiral Symmetry Restoration At Finite Temperatures In
  %SU(2) And SU(3) Gauge Theories,''
  Phys.\ Rev.\ Lett.\  {\bf 50}, 393 (1983).
  %%CITATION = PRLTA,50,393;%%

\bibitem{Fukugita:1986rr}
  M.~Fukugita and A.~Ukawa,
  %``Deconfining And Chiral Transitions Of Finite Temperature Quantum
  %Chromodynamics In The Presence Of Dynamical Quark Loops,''
  Phys.\ Rev.\ Lett.\  {\bf 57}, 503 (1986).
  %%CITATION = PRLTA,57,503;%%

\bibitem{Karsch:1994hm}
  F.~Karsch and E.~Laermann,
  %``Susceptibilities, the specific heat and a cumulant in two flavor QCD,''
  Phys.\ Rev.\  D {\bf 50}, 6954 (1994).
 % [arXiv:hep-lat/9406008].
  %%CITATION = PHRVA,D50,6954;%%

\bibitem{Digal:2000ar}
  S.~Digal, E.~Laermann and H.~Satz,
  %``Deconfinement through chiral symmetry restoration in two-flavour QCD,''
  Eur.\ Phys.\ J.\  C {\bf 18}, 583 (2001).
%  [arXiv:hep-ph/0007175].
  %%CITATION = EPHJA,C18,583;%%

\bibitem{Digal:2002wn}
  S.~Digal, E.~Laermann and H.~Satz,
  %``Interplay between chiral transition and deconfinement,''
  Nucl.\ Phys.\  A {\bf 702}, 159 (2002).
  %%CITATION = NUPHA,A702,159;%%

%%%% review %%%%%

\bibitem{Karsch:2001cy}
  F.~Karsch,
  %``Lattice QCD at high temperature and density,''
  Lect.\ Notes Phys.\  {\bf 583}, 209 (2002).
%  [arXiv:hep-lat/0106019].
  %%CITATION = LNPHA,583,209;%%

\bibitem{Laermann:2003cv}
  E.~Laermann and O.~Philipsen,
  %``Status of lattice QCD at finite temperature,''
  Ann.\ Rev.\ Nucl.\ Part.\ Sci.\  {\bf 53}, 163 (2003).
%  [arXiv:hep-ph/0303042].
  %%CITATION = ARNUA,53,163;%%

%%%% physical quark mass %%%%
%%%%%%MILC group {Bernard:2004je}, RBC-Bielefeld group {Cheng:2006qk}
%%%MILC group Bernard:2004je%%%%
%%%%% Wuppertal-Budapest group {Aoki:2006br,Aoki:2009sc,Fodor:2009ax}

\bibitem{Bernard:2004je}
  C.~Bernard {\it et al.}  [MILC Collaboration],
  %``QCD thermodynamics with three flavors of improved staggered quarks,''
  Phys.\ Rev.\  D {\bf 71}, 034504 (2005).
 % [arXiv:hep-lat/0405029].
  %%CITATION = PHRVA,D71,034504;%%

%%%%% RBC-Bielefeld group {Cheng:2006qk} %%%%
\bibitem{Cheng:2006qk}
  M.~Cheng {\it et al.},
  %``The transition temperature in QCD,''
  Phys.\ Rev.\  D {\bf 74}, 054507 (2006).
%  [arXiv:hep-lat/0608013].
  %%CITATION = PHRVA,D74,054507;%%

\bibitem{Cheng:2007jq}
  M.~Cheng {\it et al.},
  %``The QCD Equation of State with almost Physical Quark Masses,''
  Phys.\ Rev.\  D {\bf 77}, 014511 (2008)
  [arXiv:0710.0354 [hep-lat]].
  %%CITATION = PHRVA,D77,014511;%%

%%%%%hotQCD {Bazavov:2009zn,Cheng:2009zi}
\bibitem{Bazavov:2009zn}
  A.~Bazavov {\it et al.},
  %``Equation of state and QCD transition at finite temperature,''
  Phys.\ Rev.\  D {\bf 80}, 014504 (2009)
  [arXiv:0903.4379 [hep-lat]].
  %%CITATION = PHRVA,D80,014504;%%

\bibitem{Cheng:2009zi}
  M.~Cheng {\it et al.},
  %``Equation of State for physical quark masses,''
  Phys.\ Rev.\  D {\bf 81}, 054504 (2010)
  [arXiv:0911.2215 [hep-lat]].
  %%CITATION = PHRVA,D81,054504;%%

%%%{Bazavov:2009mi,Petreczky:2010tf}
\bibitem{Bazavov:2009mi}
  A.~Bazavov and P.~Petreczky  [HotQCD Collaboration],
  %``First results on QCD thermodynamics with HISQ action,''
  PoS {\bf LAT2009}, 163 (2009)
  [arXiv:0912.5421 [hep-lat]].
  %%CITATION = POSCI,LAT2009,163;%

\bibitem{Petreczky:2010tf}
  P.~Petreczky,
  %``Progress in Lattice QCD at non-zero temperature and QGP,''
 [arXiv:1012.4425 [nucl-th]].


%%%%% Wuppertal-Budapest group {Aoki:2006we,Aoki:2006br,Aoki:2009sc,
%%%%% Fodor:2009ax,Borsanyi:2010bp}%%%%%%

\bibitem{Aoki:2006we}
  Y.~Aoki, G.~Endrodi, Z.~Fodor, S.~D.~Katz and K.~K.~Szabo,
  %``The order of the quantum chromodynamics transition predicted by the
  %standard model of particle physics,''
  Nature {\bf 443}, 675 (2006)
  [arXiv:hep-lat/0611014].
  %%CITATION = NATUA,443,675;%%

\bibitem{Aoki:2006br}
  Y.~Aoki, Z.~Fodor, S.~D.~Katz and K.~K.~Szabo,
  %``The QCD transition temperature: Results with physical masses in the
  %continuum limit,''
  Phys.\ Lett.\  B {\bf 643}, 46 (2006).
 % [arXiv:hep-lat/0609068].
  %%CITATION = PHLTA,B643,46;%%

\bibitem{Aoki:2009sc}
  Y.~Aoki, S.~Borsanyi, S.~Durr, Z.~Fodor, S.~D.~Katz, S.~Krieg and K.~K.~Szabo,
  %``The QCD transition temperature: results with physical masses in the
  %continuum limit II,''
  JHEP {\bf 0906}, 088 (2009).
  % [arXiv:0903.4155 [hep-lat]].
  %%CITATION = JHEPA,0906,088;%%

\bibitem{Fodor:2009ax}
  Z.~Fodor and S.~D.~Katz,
  %``The phase diagram of quantum chromodynamics,''
  arXiv:0908.3341 [hep-ph].
  %%CITATION = ARXIV:0908.3341;%%

\bibitem{Borsanyi:2010bp}
  S.~Borsanyi, Z.~Fodor, C.~Hoelbling, S.~D.~Katz, S.~Krieg, C.~Ratti and K.~K.~Szabo
                  [Wuppertal-Budapest Collaboration],
  %``Is there still any Tc mystery in lattice QCD? Results with physical masses
  %in the continuum limit III,''
  JHEP {\bf 1009}, 073 (2010)
  [arXiv:1005.3508 [hep-lat]].
  %%CITATION = JHEPA,1009,073;%%

%%%%%%%%%%%%%%
%%%%%%%%%%%%%%
%%%PNJL%%% Fukushima:2003fw,Ratti:2005jh,Ghosh:2006qh,Fu:2007xc,
%%%Zhang:2006gu,Fukushima:2008wg
%%%PLSM%%%  Schaefer:2007pw,Mao:2009aq
\bibitem{Fukushima:2003fw}
  K.~Fukushima,
  %``Chiral effective model with the Polyakov loop,''
  Phys.\ Lett.\  B {\bf 591}, 277 (2004).
 % [arXiv:hep-ph/0310121].
  %%CITATION = PHLTA,B591,277;%%

\bibitem{Ratti:2005jh}
  C.~Ratti, M.~A.~Thaler and W.~Weise,
  %``Phases of QCD: Lattice thermodynamics and a field theoretical model,''
  Phys.\ Rev.\  D {\bf 73}, 014019 (2006).
 % [arXiv:hep-ph/0506234].
  %%CITATION = PHRVA,D73,014019;%%

\bibitem{Sasaki:2006ww}
 C.~Sasaki, B.~Friman and K.~Redlich,
  %``Susceptibilities and the phase structure of a chiral model with  Polyakov
  %loops,''
  Phys.\ Rev.\  D {\bf 75}, 074013 (2007)
  [arXiv:hep-ph/0611147].

\bibitem{Ghosh:2006qh}
  S.~K.~Ghosh, T.~K.~Mukherjee, M.~G.~Mustafa and R.~Ray,
  %``Susceptibilities and speed of sound from PNJL model,''
  Phys.\ Rev.\  D {\bf 73}, 114007 (2006).
  %[arXiv:hep-ph/0603050].
  %%CITATION = PHRVA,D73,114007;%%

\bibitem{Fu:2007xc}
W.~j.~Fu, Z.~Zhang and Y.~x.~Liu,
  %``2+1 Flavor Polyakov--Nambu--Jona-Lasinio Model at Finite Temperature and
  %Nonzero Chemical Potential,''
  Phys.\ Rev.\  D {\bf 77}, 014006 (2008).
  %[arXiv:0711.0154 [hep-ph]].
  %%CITATION = PHRVA,D77,014006;%%

\bibitem{Zhang:2006gu}
  Z.~Zhang and Y.~X.~Liu,
  %``Coupling of pion condensate, chiral condensate and Polyakov loop in an
  %extended NJL model,''
  Phys.\ Rev.\  C {\bf 75}, 064910 (2007).
 % [arXiv:hep-ph/0610221].
  %%CITATION = PHRVA,C75,064910;%%

\bibitem{Fukushima:2008wg}
  K.~Fukushima,
  %``Phase diagrams in the three-flavor Nambu--Jona-Lasinio model with the
  %Polyakov loop,''
  Phys.\ Rev.\  D {\bf 77}, 114028 (2008)
  [Erratum-ibid.\  D {\bf 78}, 039902 (2008)].
  %[arXiv:0803.3318 [hep-ph]].
  %%CITATION = PHRVA,D77,114028;%%

\bibitem{Abuki:2008nm}
  H.~Abuki, R.~Anglani, R.~Gatto, G.~Nardulli and M.~Ruggieri,
  %``Chiral crossover, deconfinement and quarkyonic matter within a Nambu-Jona
  %Lasinio model with the Polyakov loop,''
  Phys.\ Rev.\  D {\bf 78}, 034034 (2008).
 % [arXiv:0805.1509 [hep-ph]].
  %%CITATION = PHRVA,D78,034034;%%

\bibitem{Schaefer:2007pw}
  B.~J.~Schaefer, J.~M.~Pawlowski and J.~Wambach,
  %``The Phase Structure of the Polyakov--Quark-Meson Model,''
  Phys.\ Rev.\  D {\bf 76}, 074023 (2007).
 % [arXiv:0704.3234 [hep-ph]].
  %%CITATION = PHRVA,D76,074023;%%

\bibitem{Mao:2009aq}
  H.~Mao, J.~Jin and M.~Huang,
  %``Phase diagram and thermodynamics of the Polyakov linear sigma model with
  %three quark flavors,''
  J.\ Phys.\ G {\bf 37}, 035001 (2010).
  %[arXiv:0906.1324 [hep-ph]].
  %%CITATION = JPHGB,G37,035001;%%

\bibitem{Kondo:2010ts}
  K.~I.~Kondo,
  %``Toward a first-principle derivation of confinement and
  %chiral-symmetry-breaking crossover transitions in QCD,''
  Phys.\ Rev.\  D {\bf 82}, 065024 (2010)
  [arXiv:1005.0314 [hep-th]].
  %%CITATION = PHRVA,D82,065024;%%

\bibitem{Herbst:2010rf}
  T.~K.~Herbst, J.~M.~Pawlowski and B.~J.~Schaefer,
  %``The phase structure of the Polyakov--quark-meson model beyond mean field,''
  Phys.\ Lett.\  B {\bf 696}, 58 (2011)
  [arXiv:1008.0081 [hep-ph]].
  %%CITATION = PHLTA,B696,58;%%


%%%%%% ordp1-ordp3 %%%%%%%%%%
%%%%Synatschke:2007bz,Synatschke:2008yt,Bilgici:2008ui %%%

\bibitem{Synatschke:2007bz}
  F.~Synatschke, A.~Wipf and C.~Wozar,
  %``Spectral sums of the Dirac-Wilson operator and their relation to the
  %Polyakov loop,''
  Phys.\ Rev.\  D {\bf 75}, 114003 (2007).
  %[arXiv:hep-lat/0703018].
  %%CITATION = PHRVA,D75,114003;%%

\bibitem{Synatschke:2008yt}
  F.~Synatschke, A.~Wipf and K.~Langfeld,
  %``Relation between chiral symmetry breaking and confinement in YM-theories,''
  Phys.\ Rev.\  D {\bf 77}, 114018 (2008).
  %[arXiv:0803.0271 [hep-lat]].
  %%CITATION = PHRVA,D77,114018;%%

\bibitem{Bilgici:2008ui}
  E.~Bilgici and C.~Gattringer,
  %``Static quark-antiquark potential and Dirac eigenvector correlators,''
  JHEP {\bf 0805}, 030 (2008).
 % [arXiv:0803.1127 [hep-lat]].
  %%CITATION = JHEPA,0805,030;%%


%%%%spctrls1-4 %%%
%%%Gattringer:2006ci,Bruckmann:2006kx,Bilgici:2008qy
\bibitem{Gattringer:2006ci}
  C.~Gattringer,
  %``Linking confinement to spectral properties of the Dirac operator,''
  Phys.\ Rev.\ Lett.\  {\bf 97}, 032003 (2006).
 % [arXiv:hep-lat/0605018].
  %%CITATION = PRLTA,97,032003;%%

\bibitem{Bruckmann:2006kx}
  F.~Bruckmann, C.~Gattringer and C.~Hagen,
  %``Complete spectra of the Dirac operator and their relation to confinement,''
  Phys.\ Lett.\  B {\bf 647}, 56 (2007).
 % [arXiv:hep-lat/0612020].
  %%CITATION = PHLTA,B647,56;%%

\bibitem{Bilgici:2008qy}
  E.~Bilgici, F.~Bruckmann, C.~Gattringer and C.~Hagen,
  %``Dual quark condensate and dressed Polyakov loops,''
  Phys.\ Rev.\  D {\bf 77}, 094007 (2008).
%  [arXiv:0801.4051 [hep-lat]].
  %%CITATION = PHRVA,D77,094007;%%

%%%%%%%%%%%%%%%%%%
%%%%%%%%%%cbr, Banks:1979yr%%%%

%%%%%%%%dqc1-2, Bruckmann:2008br,Bilgici-thesis,Fischer:2009wc,Kashiwa:2009ki
%%%%%%%%
\bibitem{Bruckmann:2008br}
  F.~Bruckmann, C.~Hagen, E.~Bilgici and C.~Gattringer,
  %``Dressed Polyakov loops and center symmetry from Dirac spectra,''
  PoS {\bf CONFINEMENT8}, 054 (2008).
 % [arXiv:0812.2895 [hep-lat]].
  %%CITATION = POSCI,CONFINEMENT8,054;%%

\bibitem{Bilgici-thesis} E. Bilgici, PhD Thesis, University of Garz, Austria, 2009
(http://physik.uni-garz.at/itp/files/bilgici/dissertation.pdf).

\bibitem{Zhang:2010ui}
  B.~Zhang, F.~Bruckmann, C.~Gattringer, Z.~Fodor and K.~K.~Szabo,
  %``Dual condensate and QCD phase transition,''
  arXiv:1012.2314 [hep-lat].
  %%CITATION = ARXIV:1012.2314;%%

\bibitem{Fischer:2009wc}
  C.~S.~Fischer,
  %``Deconfinement phase transition and the quark condensate,''
  Phys.\ Rev.\ Lett.\  {\bf 103}, 052003 (2009).
%  [arXiv:0904.2700 [hep-ph]].
  %%CITATION = PRLTA,103,052003;%%

\bibitem{Fischer:2009gk}
  C.~S.~Fischer and J.~A.~Mueller,
  %``Chiral and deconfinement transition from Dyson-Schwinger equations,''
  Phys.\ Rev.\  D {\bf 80}, 074029 (2009).
 % [arXiv:0908.0007 [hep-ph]].
  %%CITATION = PHRVA,D80,074029;%%

\bibitem{Fischer:2010fx}
  C.~S.~Fischer, A.~Maas and J.~A.~Mueller,
  %``Chiral and deconfinement transition from correlation functions: SU(2) vs.
  %SU(3),''
  arXiv:1003.1960 [hep-ph].
  %%CITATION = ARXIV:1003.1960;%%

\bibitem{Kashiwa:2009ki}
  K.~Kashiwa, H.~Kouno and M.~Yahiro,
  %``Dual quark condensate in the Polyakov-loop extended NJL model,''
  Phys.\ Rev.\  D {\bf 80}, 117901 (2009).
 % [arXiv:0908.1213 [hep-ph]].
  %%CITATION = PHRVA,D80,117901;%%

\bibitem{Gatto:2010qs}
  R.~Gatto and M.~Ruggieri,
  %``Dressed Polyakov loop and phase diagram of hot quark matter under magnetic
  %field,''
  Phys.\ Rev.\  D {\bf 82}, 054027 (2010)
  [arXiv:1007.0790 [hep-ph]].
  %%CITATION = PHRVA,D82,054027;%%

\bibitem{Mukherjee:2010cp} T.~K.~Mukherjee, H.~Chen and M.~Huang,
  %``Chiral condensate and dressed Polyakov loop in the Nambu--Jona-Lasinio
  %model,''
  Phys.\ Rev.\  D {\bf 82}, 034015 (2010).
  %[arXiv:1005.2482 [hep-ph]].
  %%CITATION = PHRVA,D82,034015;%%

\bibitem{NJL-report}
U.~Vogl and W.~Weise,
%``The Nambu and Jona Lasinio model: Its implications for hadrons and nuclei,''
Prog.\ Part.\ Nucl.\ Phys.\  {\bf 27}, 195 (1991);
%%CITATION = PPNPD,27,195;%%
%\cite{Klevansky:1992qe}
%\bibitem{Klevansky:1992qe}
S.P.~Klevansky,
%``The Nambu-Jona-Lasinio model of quantum chromodynamics,''
Rev.\ Mod.\ Phys.\  {\bf 64}, 649 (1992);
%%CITATION = RMPHA,64,649;%%
%\cite{Hatsuda:1994pi}
%\bibitem{Hatsuda:1994pi}
T.~Hatsuda and T.~Kunihiro,
%``QCD phenomenology based on a chiral effective Lagrangian,''
Phys.\ Rept.\  {\bf 247}, 221 (1994);
%[ hep-ph/9401310].
%%CITATION = HEP-PH 9401310;%%
R.~Alkofer, H.~Reinhardt and H.~Weigel,
%``Baryons as chiral solitons in the Nambu-Jona-Lasinio model,''
Phys.\ Rept.\  {\bf 265}, 139 (1996);

\bibitem{Rehberg-Klevansky1995}
  P.~Rehberg, S.~P.~Klevansky and J.~Hufner,
  %``Hadronization in the SU(3) Nambu-Jona-Lasinio model,''
  Phys.\ Rev.\  C {\bf 53}, 410 (1996)
  [arXiv:hep-ph/9506436].
  %%CITATION = PHRVA,C53,410;%%

\bibitem{Buballa:2003qv}
M.~Buballa,
  %``NJL model analysis of quark matter at large density,''
  Phys.\ Rept.\  {\bf 407}, 205 (2005).
%  [arXiv:hep-ph/0402234].
  %%CITATION = PRPLC,407,205;%%

\bibitem{Abuki2010}
H.~Abuki, G.~Baym, T.~Hatsuda and N.~Yamamoto,
  %``The NJL model of dense three-flavor matter with axial anomaly: the low
  %temperature critical point and BEC-BCS diquark crossover,''
 % arXiv:1003.0408 [hep-ph].
 Phys.\ Rev.\  D {\bf 81}, 125010 (2010).
  %%CITATION = ARXIV:1003.0408;%%

\bibitem{Pisarski:1983ms}
  R.~D.~Pisarski, F.~Wilczek,
  %``Remarks on the Chiral Phase Transition in Chromodynamics,''
  Phys.\ Rev.\  {\bf D29}, 338-341 (1984).

\bibitem{costa-NJL}
P.~Costa, M.~C.~Ruivo and C.~A.~de Sousa,
  %``Thermodynamics and critical behavior in the Nambu-Jona-Lasinio model of
  %QCD,''
  Phys.\ Rev.\  D {\bf 77}, 096001 (2008).
%  [arXiv:0801.3417 [hep-ph]].
  %%CITATION = PHRVA,D77,096001;%%

%% {CFL,CFLK,udSC}
\bibitem{CFL} M.G.~Alford, K.~Rajagopal and F.~Wilczek,
%``Color-flavor locking and chiral symmetry breaking in ...''
Nucl.\ Phys.\ {\bf B537}, 443 (1999).
%%CITATION = HEP-PH 9804403;%%

\bibitem{CFL-K}
T.~Schafer,
%``Kaon condensation in high density quark matter,''
Phys.\ Rev.\ Lett.\  {\bf 85}, 5531 (2000);
%[ nucl-th/0007021];
%\cite{Bedaque:2001je}
%\bibitem{Bedaque:2001je}
P.F.~Bedaque and T.~Schafer,
%``High density quark matter under stress,''
Nucl.\ Phys.\ A {\bf 697}, 802 (2002);
%[ hep-ph/0105150];
%%CITATION = HEP-PH 0105150;%%
%%CITATION = NUCL-TH 0007021;%%
D.B.~Kaplan and S.~Reddy,
%``Novel phases and transitions in quark matter,''
Phys.\ Rev.\ D {\bf 65}, 054042 (2002).
%[ hep-ph/0107265].
%%CITATION = HEP-PH 0107265;%%

%\cite{Iida:2003cc}
\bibitem{udSC}
K.~Iida, T.~Matsuura, M.~Tachibana and T.~Hatsuda,
%``Melting pattern of diquark condensates in quark matter,''
Phys.\ Rev.\ Lett.\  {\bf 93}, 132001 (2004).
%% hep-ph/0312363.
%%CITATION = HEP-PH 0312363;%%

\bibitem{He:2005nk}
  L.~y.~He, M.~Jin and P.~f.~Zhuang,
  %``Pion superfluidity and meson properties at finite isospin density,''
  Phys.\ Rev.\  D {\bf 71}, 116001 (2005)
  [arXiv:hep-ph/0503272].
  %%CITATION = PHRVA,D71,116001;%%

\bibitem{g2SC}
I.~Shovkovy and M.~Huang,
%``Gapless two-flavor color superconductor,''
Phys.\ Lett.\ B {\bf 564}, 205 (2003);
 M.~Huang and I.~Shovkovy,
%``Gapless color superconductivity at zero and at finite temperature,''
Nucl.\ Phys.\ A {\bf 729}, 835 (2003);
%[ hep-ph/0307273].
%%CITATION = HEP-PH 0307273;%%
%\cite{Huang:2004bg}
%\bibitem{chromo-ins-g2SC}
M.~Huang and I.A.~Shovkovy,
%``Chromomagnetic instability in dense quark matter,''
Phys.\ Rev.\  D {\bf 70}, 051501 (2004);
%hep-ph/0407049,
%%CITATION = HEP-PH 0407049;%%
%\cite{Huang:2004bg}
%\bibitem{PME-HS}
M.~Huang and I.A.~Shovkovy,
%``Screening masses in neutral two-flavor color superconductor,''
Phys.\ Rev.\  D {\bf 70}, 094030 (2004).
%% hep-ph/0408268.
%%CITATION = HEP-PH 0408268;%%

%\cite{Alford:2003fq}
\bibitem{gCFL}
M.~Alford, C.~Kouvaris and K.~Rajagopal,
%``Gapless color-flavor-locked quark matter,''
Phys.\ Rev.\ Lett.\  {\bf 92}, 222001 (2004);
%[ hep-ph/0311286];
%%CITATION = HEP-PH 0311286;%%
%\cite{Alford:2004hz}
%\bibitem{Alford:2004hz}
M.~Alford, C.~Kouvaris and K.~Rajagopal,
%``Evaluating the gapless color-flavor locked phase,''
Phys.\ Rev.\  D {\bf 71}, 054009 (2005).
%% hep-ph/0406137.
%%CITATION = HEP-PH 0406137;%%

%% LOFF, GluonCond,current,
\bibitem{LOFF}
I.~Giannakis and H.~C.~Ren,
  %``Chromomagnetic instability and the LOFF state in a two flavor color
  %superconductor,''
  Phys.\ Lett.\ B {\bf 611}, 137 (2005);
  I.~Giannakis and H.~C.~Ren,
  %``The Meissner effect in a two flavor LOFF color superconductor,''
 Nucl.\ Phys.\ B {\bf 723}, 255 (2005);
 %hep-th/0504053.
  I.~Giannakis, D.~f.~Hou and H.~C.~Ren,
  %``A neutral two flavor LOFF color superconductor,''
  Phys.\ Lett.\ B {\bf 631}, 16 (2005);
R.~Casalbuoni and G.~Nardulli,
%``Inhomogeneous superconductivity in condensed matter and QCD,''
Rev.\ Mod.\ Phys.\  {\bf 76}, 263 (2004);
%[ hep-ph/0305069].
%%CITATION = HEP-PH 0305069;%%
%\cite{Casalbuoni:2004wm}
%\bibitem{Casalbuoni:2004wm}
R.~Casalbuoni, M.~Ciminale, M.~Mannarelli, G.~Nardulli, M.~Ruggieri
and R.~Gatto,
%``Effective gap equation for the inhomogeneous LOFF superconductive phase,''
Phys.\ Rev.\  D {\bf 70}, 054004 (2004);
%hep-ph/0404090;
%%CITATION = HEP-PH 0404090;%%
%\cite{Casalbuoni:2004zy}
%\bibitem{Casalbuoni:2004zy}
R.~Casalbuoni, R.~Gatto, M.~Mannarelli, G.~Nardulli and M.~Ruggieri,
%``Magnetic properties of the Larkin-Ovchinnikov-Fulde-Ferrell superconducting
%phase,''
Phys.\ Lett.\  B {\bf 600}, 48 (2004).
%hep-ph/0407210.
%%CITATION = HEP-PH 0407210;%%


\bibitem{GC}
 E.~V.~Gorbar, M.~Hashimoto and V.~A.~Miransky,
  %``Gluonic phase in neutral two-flavor dense QCD,''
  Phys.\ Lett.\ B {\bf 632}, 305 (2006);
M.~Hashimoto,
  %``Manifestation of Instabilities in Nambu-Jona-Lasinio type models,''
Phys.\ Lett.\  B {\bf 642}, 93 (2006).
%% hep-ph/0605323;
  %%CITATION = HEP-PH 0605323;%%

\bibitem{current}
D.~K.~Hong,
  %``RG analysis and magnetic instability in gapless superconductors,''
  hep-ph/0506097;
Mei Huang,
  %``Understanding magnetic instability in gapless superconductors,''
  Int.\ J.\ Mod.\ Phys.\ A {\bf 21}, 910 (2006);
  Mei Huang,
 %``Spontaneous current generation in the 2SC phase,''
  Phys.\ Rev.\ D {\bf 73}, 045007 (2006);
A.~Kryjevski,
  %``Spontaneous superfluid current generation in CFL at nonzero strange quark
  %mass,''
Phys.\ Rev.\  D {\bf 77}, 014018 (2008);
%%hep-ph/0508180;
  %%CITATION = HEP-PH 0508180;%%
T.~Schafer,
%``Meson supercurrent state in high density QCD,''
  Phys.\ Rev.\ Lett.\  {\bf 96}, 012305 (2006);
 % hep-ph/0508190;
 A.~Gerhold and T.~Schafer,
  %``Meson current in the CFL phase,''
 Phys.\ Rev.\  D {\bf 73}, 125022 (2006).
 %%hep-ph/0603257.

\bibitem{Higgs-current}
 I.~Giannakis, D.~Hou, M.~Huang and H.~c.~Ren,
  %``Higgs instability in gapless superfluidity / superconductor,''
  Phys.\ Rev.\  D {\bf 75}, 011501 (2007);
 %% [arXiv:hep-ph/0606178];
  %%CITATION = PHRVA,D75,011501;%%
  I.~Giannakis, D.~Hou, M.~Huang and H.~c.~Ren,
  %``Inhomogeneity driven by Higgs instability in gapless superconductor,''
  Phys.\ Rev.\  D {\bf 75}, 014015 (2007).
 %% [arXiv:hep-ph/0609098].
  %%CITATION = PHRVA,D75,014015;%%



\end{thebibliography}
\end{document}